\documentclass[twocolumn,showpacs,amsmath,amssymb,pre,floatfix]{revtex4}

\usepackage{graphicx}
\usepackage{dcolumn}
\usepackage{bm}

\begin{document}

\preprint{APS/123-QED}

\title{Mean Escape Time in a System with Stochastic Volatility}

\author{Giovanni Bonanno, Davide Valenti$^{\star}$ and Bernardo Spagnolo$^{\dagger}$}%
 \affiliation{Dipartimento di Fisica e Tecnologie Relative,
  Group of Interdisciplinary
  Physics\footnote {Electronic address: http://gip.dft.unipa.it},
  Universit\`a di Palermo,
 \\
Viale delle Scienze, pad.~18, I-90128 Palermo, Italy
\\$^{\star}$valentid@gip.dft.unipa.it,
$^{\dagger}$spagnolo@unipa.it}
\date{\today}

\begin{abstract}
We study the mean escape time in a market model with stochastic
volatility. The process followed by the volatility is the Cox
Ingersoll and Ross process which is widely used to model stock price
fluctuations. The market model can be considered as a generalization
of the Heston model, where the geometric Brownian motion is replaced
by a random walk in the presence of a cubic nonlinearity. We
investigate the statistical properties of the escape time of the
returns, from a given interval, as a function of the three
parameters of the model. We find that the noise can have a
stabilizing effect on the system, as long as the global noise is not
too high with respect to the effective potential barrier experienced
by a fictitious Brownian particle. We compare the probability
density function of the return escape times of the model with those
obtained from real market data. We find that they fit very well.
\end{abstract}

\maketitle

\section{\label{sec:intro}Introduction}
The noise in physical system gives rise to interesting and sometimes
counterintuitive effects. The stochastic resonance and the noise
enhanced stability are two examples of noise activated phenomena
that have been extensively studied in a wide variety of natural and
physical systems such as lasers, spin systems, chemical and
biological complex systems~\cite{SR,NES,Nes-theory}. Specifically
the activated escape from a metastable state is important in the
description of the dynamics of non-equilibrium complex
systems~\cite{metastable1,metastable2}. Recently there has been a
growing interest in the application of complex systems methodology
to model social systems. In particular the application of
statistical physics for modeling the behavior of financial markets
has given rise to a new field called {\it
econophysics}~\cite{Complex, Man-Stan, Bouch-Pott,Cont}.

The stock price evolution is indeed driven by the interaction of a
great number of traders. Each one follows his own strategy in order
to maximize his profit. There are fundamental traders who try to
invest in solid company, speculators who ever try to exploit
arbitrage opportunity and also noise traders who act in a
non-rational way. All these considerations allow us to say that the
market can be thought as a complex system where the rationality and
the arbitrariness of human decisions are modeled by using stochastic
processes.

The price of financial time series was modeled as a random walk, for
the first time, by Bachelier~\cite{Bachelier}. His model provides
only a rough approximation of the real behavior of the price time
series. Indeed it doesn't reproduce some of the stylized facts of
the financial markets: (i) the distribution of relative price
variation (price return) has fat tails, showing strong
non-Gaussianity~\cite{Man-Stan,Cont}; (ii) the standard deviation of
the return time series, called volatility, is a stochastic process
itself characterized by long memory and
clustering~\cite{Dacorogna,Cont}; (iii) autocorrelations of asset
returns are often negligible~\cite{Cont}.

A popular model proposed to characterize the stochastic nature of
the volatility is the Heston model~\cite{Heston}, where the
volatility is described by a process known as the Cox, Ingersoll and
Ross (CIR) process ~\cite{Cir-Fouque} and in mathematical statistics
as the Feller process~\cite{Feller}. The model has been recently
investigated by econophysicists~\cite{Silva,BonannoFNL} and solved
analytically~\cite{Micci,Yakovenko}. Models of financial markets
reproducing the most prominent features of statistical properties of
stock market fluctuations and whose dynamics is governed by
non-linear stochastic differential equations have been considered
recently in
literature~\cite{Solomon,Borland,Borland1,Kuhn,BouchaudCont,Bouchaud01,Bouchaud02,Sornette}.
Moreover financial markets present days of normal activity and
extreme days of crashes and rallies characterized by different
behaviors for the volatility. The question whether extreme days are
outliers or not is still debated. This research topic has been
addressed both by physicists~\cite{Sornette} and
economists~\cite{Friedman}.

A Langevin approach to the market dynamics, where market crisis was
modeled through the use of a cubic potential with a metastable
state, was already
proposed~\cite{Bouch-Pott,BouchaudCont,Bouchaud01,Bouchaud02}. There
feedback effects on the price fluctuations were considered in a
stochastic dynamical equation for instantaneous returns. The
evolution inside the metastable state represents the normal market
behavior, while the escape from the metastable state represents the
beginning of a crisis.

Systems with metastable states are ubiquitous in physics. Such
systems have been extensively studied. In particular it has been
proven that the noise can have a stabilizing effect on these
systems~\cite{NES,Nes-theory}. To the best of our knowledge all
models proposed up to now to study the escape from a metastable
state contain only a constant noise intensity, which represents in
econophysics the volatility. Recently theoretical and empirical
investigations have been done on the mean exit time (MET) of
financial time series~\cite{Mas,Mon}, that is the mean time when the
stochastic process leaves, for the first time, a given interval. The
authors investigated the MET of asset prices outside a given
interval of size $L$, and they found that the MET follows a
quadratic growth in terms of the region size $L$. Their theoretical
investigation was done within the formalism of the continuous time
random walk. Within the same formalism the statistical properties of
the waiting times for high-frequency financial data have been
investigated in refs.~\cite{Scalas}.

In this work we model the volatility with the CIR process and
investigate the statistical properties of the escape times when both
an effective potential with a metastable state and the CIR
stochastic volatility are present. Our study provides a natural
evolution of the models with constant volatility. The analysis has
the purpose to investigate the role of the noise in financial market
extending a popular market model, and to provide also a starting
model for physical systems under the influence of a fluctuating
noise intensity. The paper is organized as follows. In the next
section the modified Heston model and the noise enhanced stability
effect are described. In the third section the results for two
extreme cases of this model are reported. In section $IV$ we comment
the results for the general case and the probability density
function of the escape time of the returns, obtained by our model,
is compared with that extracted from experimental data of a real
market. In the final section we draw our conclusions.

\section{\label{sec:nes_heston} The modified Heston model and the NES effect}

The Heston model, which describes the dynamics of stock prices
$p(t)$ as a geometric Brownian motion with the volatility given by
the CIR mean-reverting process, is defined by the following Ito
stochastic differential equations

\begin{eqnarray}
dp(t) &=& \mu~p~dt + \sigma(t)p~dW_1(t) \\
dv(t) &=& a(b - v(t))~dt + c~\sqrt{v(t)}~dW_2(t),
\label{heston eq}
\end{eqnarray}
where $\sigma(t)$ is the time-dependent volatility, $ v(t) =
\sigma^2(t)$ and $W_i(t)$ are uncorrelated Wiener processes with the
usual statistical properties

\begin{equation}
<dW_i> = 0, \; <dW_i(t) dW_j(t')>\;=\; dt\;
\delta(t-t')\;\delta_{i,j}.
\end{equation}
In Eq.~(1) $\mu$ represents  a drift at macroeconomic scales. In
Eq.~(\ref{heston eq}) the volatility $\sigma(t) = \sqrt{v(t)}$
reverts towards a macroeconomic long time term given by the mean
squared value $b$, with a relaxation time $a^{-1}$. Here $c$ is the
amplitude of volatility fluctuations often called the
\emph{volatility of volatility}. By introducing log-returns $x(t) =
ln(p(t)/p(0))$ in a time window $[0,t]$ and using It$\hat{o}$'s
formula~\cite{Gardiner}, we obtain the stochastic differential
equation (SDE) for $x(t)$

\begin{equation}
  dx(t) = (\mu - v(t)/2) ~dt + \sqrt{v(t)} ~dW_1(t).
\end{equation}

\begin{figure}[htbp]
\vspace{5mm}
\centering{\resizebox{8.5cm}{!}{\includegraphics{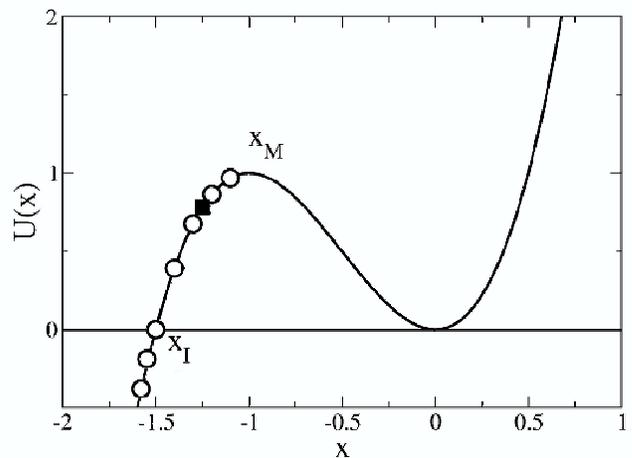}}}
\vskip-0.2cm \caption{\label{cubic_pot} Cubic potential used in the
dynamical equation for the variable $x(t)$. The points evidenced in
the figure indicate the starting positions $x_0$ used for the
simulations illustrated in section \ref{sec:limit} (white circles)
and section~\ref{sec:generic} (black square).}
\end{figure}

Here we consider a generalization of the Heston model, by replacing
the geometric Brownian motion with a random walk in the presence of
a cubic nonlinearity. This generalization represents a fictitious
"\emph{Brownian particle}" moving in an \emph{effective} potential
with a metastable state, in order to model those systems with two
different dynamical regimes like financial markets in normal
activity and extreme days~\cite{Bouchaud02,BouchaudCont,Bouchaud01}.
The equations of the new model are

\begin{eqnarray}
  dx(t)  &=&  - \left(\frac{\partial U}{\partial x} +
  \frac{v(t)}{2}\right)~dt + \sqrt{v(t)} ~dW_1(t) \\
  dv(t)  &=&  a(b-v(t)) ~dt + c \sqrt{v(t)} ~dW_2(t),
\label{Eqn:BS}
\end{eqnarray}
where $U(x)=Ax^3+Bx^2$ is the \emph{effective} cubic potential shown
in Fig.~\ref{cubic_pot} with $A=2$ and $B=3$. From now on we call
the first and second term of Eq.~(\ref{Eqn:BS}) reverting term and
noise term respectively.

Let us call $x_M$ the abscissa of the potential maximum and $x_I$
the cross point between the potential and the $x$ axes. The
intervals $x_o < x_I$ and $I=[x_I,x_M]$ are clearly regions of
instability for the system. In systems with a metastable state, the
random fluctuations can originate the noise enhanced stability (NES)
phenomenon, an interesting effect that increases the stability,
enhancing the lifetime of the metastable
state~\cite{NES,Nes-theory}. The mean escape time $\tau$ for a
Brownian particle moving throughout a barrier $\Delta U$ is given by
the the well known exponential Kramers law~\cite{Gardiner,Hanggi}

\begin{equation}
  \tau = C exp\left[{\frac{\Delta U}{v}}\right],
\label{Eqn:Kramers}
\end{equation}
where $\tau$ is a monotonically decreasing function of the noise
intensity $v$, and $C$ is a prefactor which depends on the potential
profile. This is true only if the random walk starts from initial
positions inside the potential well. When the starting position is
chosen in the instability region $x_o < x_M$, $\tau$ exhibits an
enhancement behavior, with respect to the deterministic escape time,
as a function of $v$. This is the NES effect and it can be explained
by considering the barrier "\emph{seen}" by the Brownian particle
starting at the initial position $x_0 $, that is $\Delta U_{in} =
U(x_{max})-U(x_0)$. Moreover $\Delta U_{in}$ is less than $\Delta U$
as long as the starting position $x_0$ lies into the interval
$I=[x_I,x_M]$. Therefore for a Brownian particle starting from an
unstable initial position it is more likely to enter into the well
than to escape from, once the particle has entered. So a small
amount of noise can increase the lifetime of the metastable state.
For a detailed discussion on this point and different dynamical
regimes see Refs.~\cite{Nes-theory}. When the noise intensity $v$ is
much greater than $\Delta U$, the
Kramers behavior is recovered.\\
\indent We have already investigated the statistical properties of
the escape times for models with stochastic volatility, namely the
Heston model and the GARCH model~\cite{BonannoFNL}. We found that
the probability density function of the escape times obtained in the
Heston model exhibits a better agreement with real data than that
calculated in the GARCH model. Here, by considering the modified
Heston model (Eqs.~(5) and~(\ref{Eqn:BS})), characterized by a
stochastic volatility and a nonlinear Langevin equation for the
returns, we study the mean escape time as a function of the model
parameters $a$, $b$ and $c$. In particular we investigate whether it
is possible to observe some kind of nonmonotonic behavior such that
observed for $\tau~vs.~v$ in the NES effect with constant volatility
$v$. We find a nonmonotonic behavior of the mean escape time (MET),
$\tau$, as a function of the model parameters. Within this behaviour
we recognize the enhancement of $\tau$ as a NES effect in a broad sense.\\
 \indent Our modified Heston model has
two limit regimes, corresponding to the cases $a=0$, with only the
noise term in the equation for the volatility $v(t)$, and $c=0$ with
only the reverting term in the same equation. This last case
corresponds to the usual parametric constant volatility regime. In
fact, apart from an exponential transient, the volatility reaches
the asymptotic value $b$. The NES effect should be observable in the
latter case as a function of $b$, which is the average volatility.
In fact, in this case we recover the motion of a Brownian particle
in a fixed cubic potential with a metastable state and an
enhancement of its lifetime for particular initial conditions.\\
\indent For this purpose we perform simulations by integrating
numerically the equations~(5) and~(\ref{Eqn:BS}) using a time step
$\Delta t=0.01$, and, as potential parameters, $A=2.0$ and $B=3.0$.
All the integration steps yielding negative values for $v$ were
rejected and repeated. The simulations were performed placing the
walker in the initial positions $x_0$ located in the unstable region
$[x_I,x_M]$ (see Fig.~\ref{cubic_pot}) and using an absorbing
barrier at $x=-6.0$. Each time the walker hits the barrier, the
escape time is registered and another simulation starts, placing the
walker at the same initial position $x_0$, but using the volatility
value of the barrier hitting time.

\section{\label{sec:limit}Limit cases}

First of all we present the result obtained in the limit cases where
we have only one of the two terms in the CIR
equation~(\ref{Eqn:BS}). Namely: (a) only the reverting term
($a\neq0, b\neq0, c=0$, revert-only case), and (b) only the noise
term ($a=0$, $b$ whatever, $c\neq0$, noise-only case) are present.
In the case (a) the volatility is practically constant and equal to
$b$, apart from an exponential transient that is negligible for
times $t \gg a^{-1}$.

\begin{figure}[htbp]
\vspace{5mm}
\centering{\resizebox{8.5cm}{!}{\includegraphics{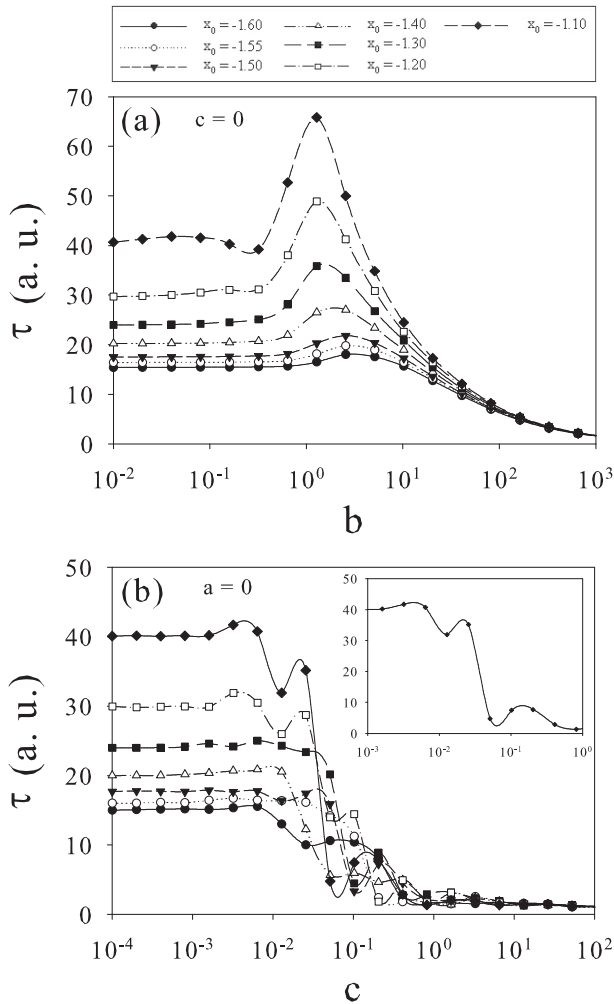}}}
\vskip-0.2cm \caption{\label{limit} Mean escape time $\tau$ for the
seven different starting positions of Fig.~\ref{cubic_pot} (white
circles in that figure). Limit cases where only one of the two terms
in the CIR equation is present: (a) $\tau$~vs.~$b$, when only the
reverting term is present ($a = 10^{-2}$, $c = 0$), and (b)
$\tau$~vs.~$c$, when only the noise term is present ($a = 0$, $b =
10^{-2}$). The different starting positions $x_0$ from top to bottom
are: $-1.1$, $-1.2$, $-1.3$, $-1.4$, $-1.5$, $-1.55$, $-1.60$. The
parameters $b$ and $c$ are dimensionless, while $a$, measured in
arbitrary units, has dimension of the inverse time. Inset: detail of
the curve with $x_0 = -1.10$.}
\end{figure}

The mean escape time as a function of $b$ is plotted in
Fig.~\ref{limit}a for the seven different initial positions
indicated in Fig.~\ref{cubic_pot} (white circles in that figure).
The curves are averaged over $10^5$ escape events. The nonmonotonic
behavior is present. The escape time increases by increasing $b$
until it reaches a maximum. After the maximum, when the values of
$b$ are much greater than the potential barrier height, the Kramers
behavior is recovered. The nonmonotonic behavior is more evident for
starting positions near the maximum of the potential. For starting
positions $x_0$ lying in the interval $I=[x_I,x_M]$, the initial
plateau is an artifact of the calculus. Indeed in the theoretical
solution for the constant volatility case, $\tau$ diverges as the
noise intensity approaches zero~\cite{NES,Nes-theory}. With such a
noise intensity the escape from the well is a very unlikely event.
We should require an infinite number of simulation steps in order to
observe an event pushing the particle into the well.

In the case (b), $v$ is a multiplicative stochastic process with
standard deviation equal to $c$. The random term in
Eq.~(\ref{Eqn:BS}) is given by the product of the white noise and
the square root of the process itself $v$. In Fig.~\ref{limit}b we
plot $\tau$ as a function of the parameter $c$. The nonmonotonic
behavior is absent and the MET decreases monotonically, but
differently from the Kramers behavior. The curves are averaged over
$10^8$ escape events. In this case a greater number of events is
required to eliminate the fluctuations present in the curves. Once
again we cannot draw any conclusions for the constant behavior of
$\tau$, when the values of the parameter $c$ are very small, because
of the finite number of simulation steps~\cite{Nes-theory}. It is
worthwhile pointing out that the NES effect is not observable as a
function of the parameter $c$, if $a = 0$. Therefore, the presence
of the reverting term affects the behavior of $\tau$ in the domain
of the noise term of the volatility and it regulates the transition
from nonmonotonic to monotonic regimes of MET. Moreover in this
noise-only regime the volatility is proportional to the square of
the Wiener process and therefore the fluctuations during the
monotonic decreasing behavior of $\tau$ are very large. We see three
different dynamical regimes: (i) for low values of the parameter
$c$, with respect to the height of the potential barrier, we have a
constant behavior, (ii) for intermediate values of $c$ we have
fluctuations with decreasing monotonic behavior, (iii) for values of
$c$ greater than $1$, which is exactly the height of the barrier, we
get very small and constant values of $\tau$ close to zero. This
behavior is mainly due to the presence of the Ito term in Eq.~(5)
for the log-returns $x(t)$. In fact, because of this term we have a
fluctuating potential, obtained by the previous one by adding a
linear term: $U^{'}= U(x)+ (xv)/2$. The effect of the positive
linear term is to modify randomly the potential shape in such a way
that the potential barrier disappears for greater values of the
volatility $v(t)$. Specifically for values of $v < 3$ the potential
barrier is always present, while for $v > 3$  it disappears. This
produces a random enhancement of the escape process with a
consequent decreasing behavior of the average escape time.

\section{\label{sec:generic}Generic case}

In order to present our results for the generic case, where both the
reverting and noise terms of the CIR equation are present, we focus
on a single starting position, $x_0=-1.25$, which is located in the
middle of the interval $[x_I,x_M]$. We analyze the escape time
through the barrier using different values for parameters $a$, $b$
and $c$. We note that the average escape time $\tau$ is measured in
arbitrary units (a. u.), the parameter $a$ (measured in a.u. too)
has the dimension of the inverse time, while $b$ and $c$ are
dimensionless.
\begin{figure}[htbp]
\vspace{1mm}
\centering{\resizebox{8.5cm}{!}{\includegraphics{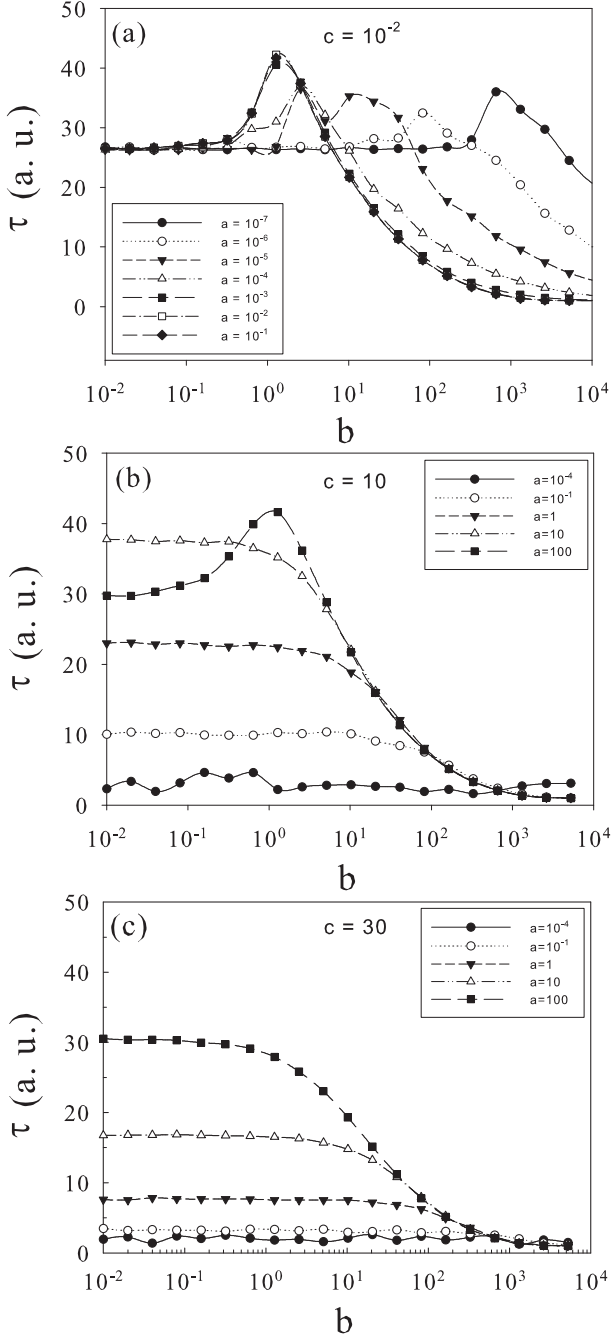}}}
 \vskip 0.5 cm \caption{\label{versusb} Mean escape time $\tau$ as a
function of reverting level $b$. Each panel corresponds to a
different value of $c$, specifically (a) $c=10^{-2}$, (b) $c = 10$
and (c) $c=30$. Inside each panel different curves correspond to the
following values of $a$: (a) black circle $10^{-7}$, white circle
$10^{-6}$, black triangle down $10^{-5}$, white triangle up
$10^{-4}$, black square $10^{-3}$, white square $10^{-2}$, black
diamond $10^{-1}$; (b) and (c) black circle $10^{-4}$, white circle
$10^{-1}$, black triangle down $1$, white triangle up $10$, black
square $10^{2}$.}
\end{figure}

As a first result we present the behavior observed for $\tau$ as a
function of the reverting level $b$. In Fig.~\ref{versusb} we show
the curves averaged over $10^5$ escape events. Each panel
corresponds to a different value of $c$. Inside each panel different
curves correspond to different values of $a$ spanning seven orders
of magnitude. The nonmonotonic shape, characteristic of the NES
effect, is clearly shown in Fig.~\ref{versusb}a. This behavior is
shifted towards higher values of $b$ as the parameter $a$ decreases,
and it is always present. In Fig.~\ref{versusb}c, which corresponds
to a much greater value of $c$ ($c = 30$), all the curves are
monotonic but with a large plateau. So an increase in the value of
$c$ causes the NES effect to disappear.

To understand this behavior let us note that the parameters $a$ and
$c$ play a regulatory role in Eq.~(\ref{Eqn:BS}). For $a \gg c$ the
drift term is predominant while for $a \ll c$ the dynamics is driven
by the noise term, unless the parameter $b$ takes great values. In
fact in Fig.~\ref{versusb}a the nonmonotonic behavior is observed
for $a \ll c$, provided that $b \gg c$. For increasing values of $a$
the system approaches the revert-only regime and we recover the
behavior shown in Fig.~\ref{limit}a. For $a \ll c$ the shape of the
curves changes. The mean escape time $\tau$ is almost constant, and
only for very high values of $b$ we observe a decreasing behavior.
This happens because for smaller values of $a$ the reverting term
becomes negligible in comparison with the noise term and the
dependence on $b$ becomes weaker. Only when $b$ is large enough the
reverting term assumes values that are no more negligible with
respect to the noise term and we can observe again a dependence of
$\tau$ on $b$. By increasing the value of $c$ we observe the
nonmonotonic behavior only for a very great value of the parameter
$a$, that is for $a = 100 \gg c$ (see Fig.~\ref{versusb}b). For
further increase of parameter $c$ (see Fig.~\ref{versusb}c), the
noise experienced by the system is much greater than the effective
potential barrier "\emph{seen}" by the fictitious Brownian particle
and the NES effect is never observable. Moreover we are near a
noise-only regime, and we can say that the magnitude of $c$ is so
high as to saturate the system.

In summary the NES effect can be observed as a function of the
volatility reverting level $b$, the effect being modulated by the
parameter $(ab)/c$. The phenomenon disappears if the noise term is
predominant in comparison with the reverting term. Moreover the
effect is no more observable if the parameter $c$ pushes the system
towards a too noisy region.

 As a second step we study the dependence of $\tau$
on the noise intensity $c$.
\begin{figure}[htbp]
\vspace{1mm}
\centering{\resizebox{8.5cm}{!}{\includegraphics{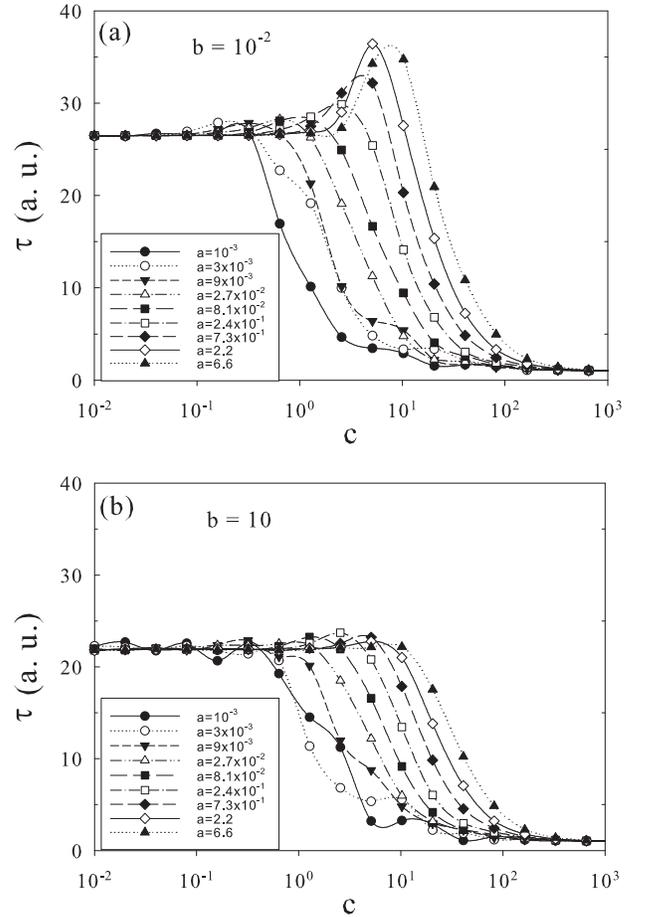}}}
\vskip-0.1cm \caption{\label{versusc} Mean escape time $\tau$ as a
function of the noise intensity $c$. Each panel corresponds to a
different value of $b$, specifically (a) $b=10^{-2}$ and (b) $b=10$.
Inside each panel different curves correspond to the following
values of $a$: black circle $10^{-3}$, white circle $3 \times
10^{-3}$, black triangle down $9 \times 10^{-3}$, white triangle up
$2.7 \times 10^{-2}$, black square $8.1 \times 10^{-2}$, white
square $2.4 \times 10^{-1}$, black diamond $7.3 \times 10^{-1}$,
white diamond $2.2$, black triangle up $6.6$. }
\end{figure}
Fig.~\ref{versusc} shows the curves of $\tau~vs~c$, averaged over
$10^5$ escape events. Each panel corresponds to a different value of
$b$. Inside each panel different curves correspond to different
values of $a$. The shape of the curves is similar to that observed
in Fig.~\ref{versusb}. For small $b$ (panel a) we observe a
nonmonotonic behavior, while for great $b$ (panel b) the curves are
monotonic but with a large plateau. Let us recall the results of
Fig.~\ref{limit}b: there was no NES effect in the noise-only case.
So for small values of $a$, when the reverting term is negligible,
the absence of the nonmonotonic behavior is expected. By increasing
$a$ the nonmonotonic behavior is recovered (see
Fig.~\ref{versusc}a). Once again, if one of the parameter pushes the
system into a high noise region, the nonmonotonic behavior
disappears (see Fig.~\ref{versusc}b). Specifically if $b$ is high,
the reverting term drives the system towards values of volatility
that are outside the region where the NES effect is observable.
Indeed a direct inspection of Fig.~\ref{versusb} shows that the
value of $b$ used in Fig.~\ref{versusc}b is located after the
maximum of $\tau$ for all values of $a$ and $c$.

 In summary when the noise term is
coupled to the reverting term we observe the NES effect on the
variable $c$. The effect disappears if $b$ is so high as to saturate
the system.
\begin{figure}[htbp]
\vspace{1mm}
\centering{\resizebox{8.5cm}{!}{\includegraphics{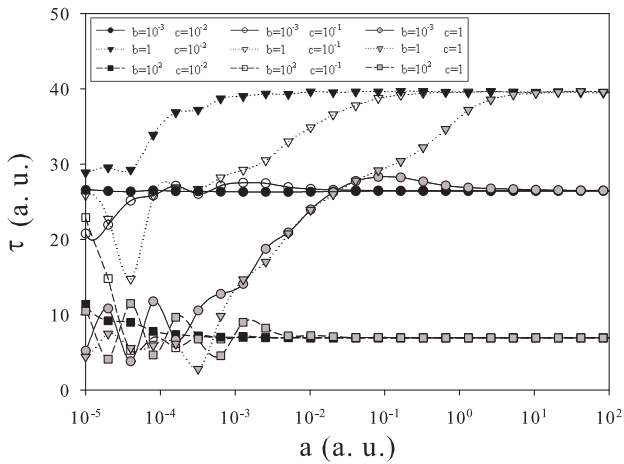}}}
\vskip-0.2cm \caption{\label{versusa} Mean escape time $\tau$ as a
function of the reverting rate $a$ for different values of $b$ and
$c$. Curves  with the same color correspond to the same value of
$c$, while curves with the same symbol correspond to the same value
of $b$. Specifically the values of $b$ used are: $10^{-3}$ circle,
$1$ triangle, $10^{2}$ square . The values of $c$ used are:
$10^{-2}$ black, $10^{-1}$ white, $1$ gray.} \label{versusa}
\end{figure}
As a last result we discuss the behavior observed for $\tau$ as a
function of the reverting rate $a$. This allow us to observe the
transition between the two regimes of the process discussed above:
the noise-only regime and the revert-only regime. The results are
reported in Fig.~\ref{versusa} for three different values of $b$ and
three different values of $c$. To reduce the fluctuations in all the
curves when the parameter $a$ becomes small, we performed
simulations by averaging on $10^6$ escape events. It is worthwhile
to note that for values of the parameter $a < 10^{-5}$, we enter in
the noise-only regime, which characterizes one of the limit cases
discussed in section III. Curves with the same color correspond to
the same value of $c$ while curves with the same symbol correspond
to the same value of $b$.

The system tends to the noise-only regime for lower values of $a$
and to the revert-only regime for higher values of $a$. On the right
end of Fig.~\ref{versusa} the curves corresponding to the same value
of $b$ tend to group together. The values of $\tau$ the curves
approach reflect the nonmonotonic behavior observed in
Fig.~\ref{versusb}a. Indeed all the curves corresponding to the
intermediate value of $b$ ($b = 1$) approach a value of $\tau$,
which almost corresponds to the maximum value of MET in
Fig.~\ref{versusb}a (we note that the behavior for $a = 10^{-1}$
coincides with that for $a = 1$, even if it is not reported in
Fig.~\ref{versusb}a). This value of $\tau$ is greater than that
reached by the curves corresponding to other two, lower and greater,
values of $b$ ($10^{-3}$ and $10^{2}$ respectively). Conversely on
the left end the curves corresponding to the same value of $c$ tend
to group together. It is worth noting that in this last case the
curves with the highest value of $c$, namely $c = 1$, show greater
fluctuations as those observed in all the previous cases where the
noise term is predominant (see for example Fig.~\ref{limit}b).

\begin{figure}[htbp]
\centering{\resizebox{8.5cm}{!}{\includegraphics{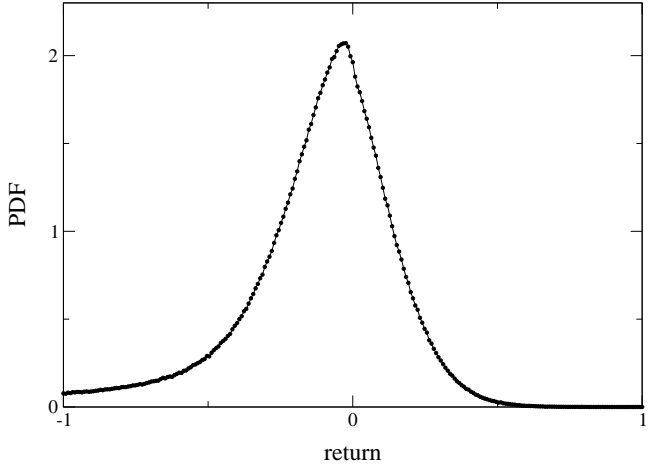}}}
\vskip-0.2cm \caption{Probability density function of the stock
price returns for our model (Eqs.~(5) and (6)). The values of the
parameters are: $a=10^{-1}$, $b=4.5$, $c=2 \times 10^{-1}$. The
potential parameters are: $A = 2$ and $B = 3$.} \label{pdf return}
\end{figure}

It is interesting to show, for our model (Eqs.~(5) and~(6)), some of
the well-established statistical signatures of the financial time
series, such as the probability density function (PDF) of the stock
price returns, the PDF of the volatility and the return correlation.
In Fig.~\ref{pdf return} we show the PDF of the returns. To
characterize quantitatively this PDF with regard to the width, the
asymmetry and the fatness of the distribution, we calculate the mean
value $<\Delta x>$, the variance $\sigma_{\Delta x}$, the skewness
$\kappa_3$, and the kurtosis $\kappa_4$. We obtain the following
values: $<\Delta x> = - 0.162$, $\sigma_{\Delta x} = 0.348$,
$\kappa_3 = - 1.958$, $\kappa_4 = 5.374$. These statistical
quantities clearly show the asymmetry of the distribution and its
leptokurtic nature observed in real market data. In fact, the
empirical PDF is characterized by a narrow and large maximum, and
fat tails in comparison with the Gaussian
distribution~\cite{Man-Stan,Bouch-Pott}. Specifically we note that
the value of the kurtosis $\kappa_4 = 5.374$, which gives a measure
of the distance between our distribution and a Gaussian one, is of
the same order of magnitude of that obtained for $S\&P$~$500$ for
daily prices (see Fig.7.2 on page 114 of Ref.~\cite{Bouch-Pott}).
The presence of the asymmetry is very interesting and it will be
subject of future investigations. This asymmetry is due to the
nonlinearity introduced in the model through the cubic potential
(see Fig.~\ref{cubic_pot}). Of course a comparison between the PDF
of real data and that obtained from the model requires further
investigations on the dynamical behavior of the system, as a
function of the model parameters. In the following Fig.~\ref{pdf
volatility} we show the PDF of the volatility for our model, and we
can see a log-normal behavior as that observed approximately in real
market data.
\begin{figure}[htbp]
\vspace{1mm}
\centering{\resizebox{8.5cm}{!}{\includegraphics{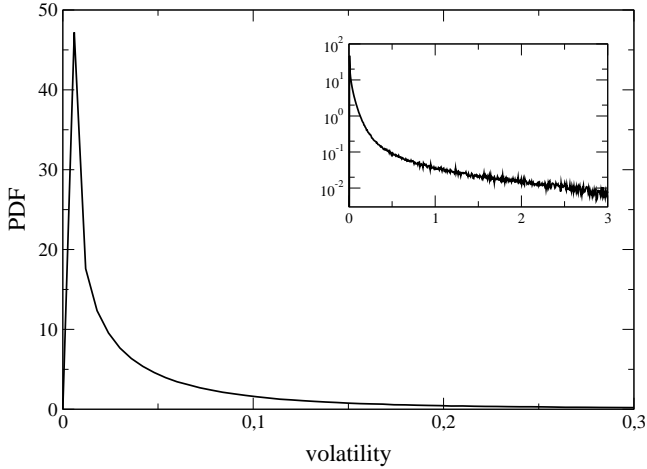}}}
\vskip-0.2cm \caption{Probability density function of the volatility
for our model (Eqs.~(5) and (6)). The values of the parameters are
the same of Fig.~\ref{pdf return}. Inset: semilog plot of the PDF of
volatility in a longer time scale.} \label{pdf volatility}
\end{figure}
Finally in Fig.~\ref{return correlation} we show the correlation
function of the returns. As we can see the autocorrelations of the
asset returns are insignificant, except for very small time scale
for which microstructure effects come into play. This is in
agreement with one of the stylized empirical facts emerging from the
statistical analysis of price variations in various types of
financial markets~\cite{Cont}. A quantitative agreement of the PDF
of volatility and the correlation of returns with the corresponding
quantities obtained from real market data is subject of further
studies.
\begin{figure}[htbp]
\vspace{1mm}
\centering{\resizebox{8.5cm}{!}{\includegraphics{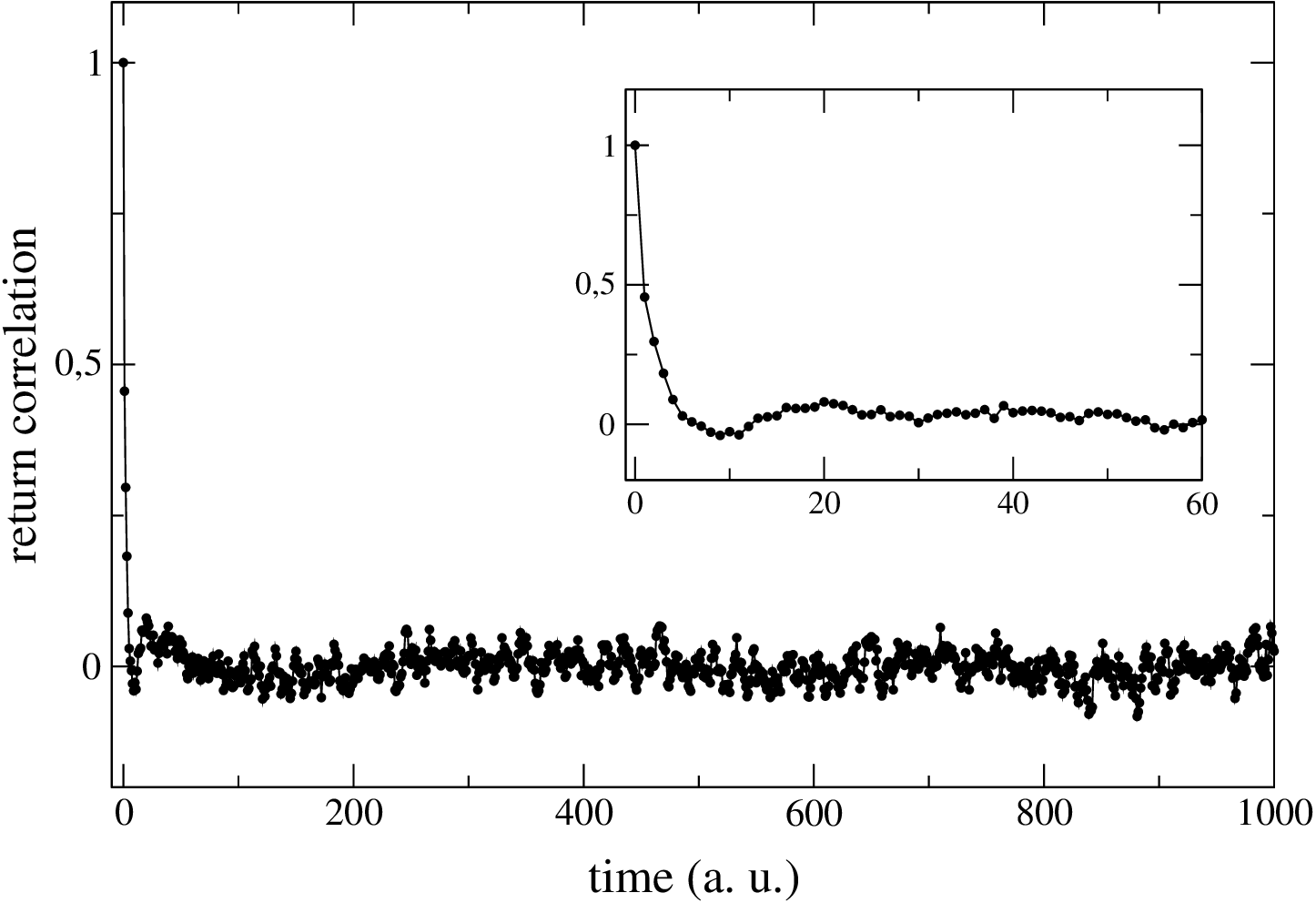}}}
\vskip-0.2cm \caption{Correlation function of the returns for our
model (Eqs.~(5) and (6)). The values of the parameters are the same
of Fig.~\ref{pdf return}. Inset: detail of the behaviour at short
times.} \label{return correlation}
\end{figure}

Our last investigation concerns the PDF of the escape time of the
returns, which is the main focus of our paper.
\begin{figure}[htbp]
\vspace{1mm}
\centering{\resizebox{8.5cm}{!}{\includegraphics{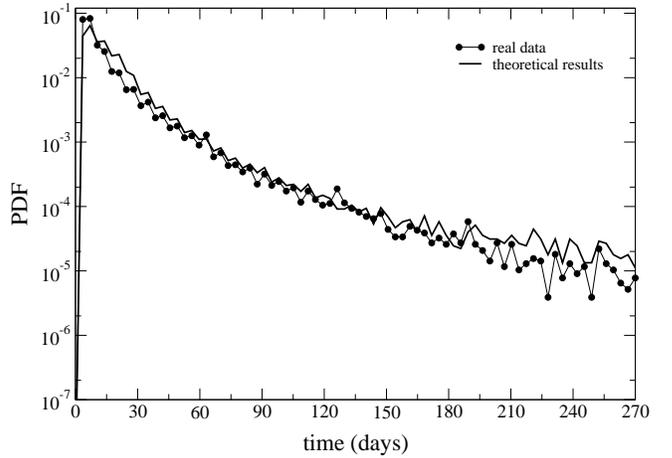}}}
\vskip-0.2cm \caption{Probability density function of the escape
time of the returns from simulation (solid line), and from real data
(black circle). The values of the parameters are: $\Delta x_i = -0.1
\thinspace \sigma_{\Delta x}$, $\Delta x_f = -1.0 \thinspace
\sigma_{\Delta x}$, $x_0 = -1.25$, $x_{abs} = -6.0$, $v_{start}=8.62
\times 10^{-5}$, $a=10^{-1}$, $b=4.5$, $c=2 \times 10^{-1}$. The
potential parameters are: $A = 2$ and $B = 3$.} \label{PDF}
\end{figure}
By using our model (Eqs.~(5) and~(\ref{Eqn:BS})), we calculate the
probability density function for the escape time of the returns. We
define two thresholds, $\Delta x_i$ and $\Delta x_f$, which
represent the start point and the end point for calculating $\tau$
respectively. When the return series reaches the value $\Delta x_i$,
the simulation starts to count the time $\tau$ and it stops when the
threshold $\Delta x_f$ is crossed. In order to fix the values of the
two thresholds we consider the standard deviation (SD)
$\sigma_{\Delta x}$ of the return series over a long time period
corresponding to that of the real data. Specifically $\sigma_{\Delta
x}$ is the average of the standard deviations $\sigma_n$ observed
for each stock during the above mentioned whole time period (n is
the stock index, varying between $1$ and $1071$). Then we set
$\Delta x_i = -0.1~\sigma_{\Delta x}$ and $\Delta x_f =
-1.0~\sigma_{\Delta x}$. We perform our simulations obtaining a
number of time series of the returns equal to the number of stocks
considered, which is $1071$. The initial position is $x_0 = -1.25$
and the absorbing barrier is at $x_{abs} = -6.0$. For the CIR
stochastic process $v$, we choose $v_{start}=8.62 \times 10^{-5}$,
$a=10^{-1}$, $b=4.5$ and $c=2 \times 10^{-1}$. The choice of this
parameter data set is not based on a fitting procedure as that used
for example in Ref.~\cite{Yakovenko}. There the minimization of the
mean square deviation between the PDF of the returns, extracted from
financial data, and that obtained theoretically is done. We choose
the parameter set in the range in which we observe the nonmonotonic
behaviour of the mean escape time of the price returns as a function
of the parameters $b$ and $c$. Then by a trial and error procedure
we select the values of the parameters $a$, $b$, and $c$ for which
we obtain the best fitting between the PDF of the escape times
calculated from the modified Heston model (Eqs. (5) and (6)) and
that obtained from the time series of real market data. We report
the results in Fig.~\ref{PDF}. Of course a better quantitative
fitting procedure could be done, by considering also the potential
parameters. This detailed analysis will be done in a forthcoming
paper.

As real data we use the daily closure prices for $1071$ stocks
traded at the NYSE and continuously present in the $12-$year period
$1987-1998$ (3030 trading days). The same data set was used in
previous investigations
\cite{BonannoFNL,Micci,Bon_PRE_03,Bon_EPJB04}. From this data set we
obtain the time series of the returns and we calculate the time to
hit a fixed threshold starting from a fixed initial position. The
two thresholds were chosen as a fraction of the average standard
deviation $\sigma_{\Delta x}$ on the whole time period, as we have
done in simulations. The agreement between real data and those
obtained from our model is very good. We note that at high escape
times the statistical accuracy is worse because of few data with
high values. The parameter values of the CIR process for which we
obtain good agreement between real and theoretical data are in the
range in which we observe the nonmonotonic behavior of MET (see
Fig.~\ref{versusb}a). This means that in this parameter region we
observe a stabilizing effect of the noise on the prices in the time
windows for which we have a variation of returns between the two
fixed values $\Delta x_i$ and $\Delta x_f$. This encourages us to
extend our analysis to large amounts of financial data and to
explore other parameter regions of the model.

\section{\label{sec:conclusions}Conclusions}

We studied the mean escape time in a market model with a cubic
nonlinearity coupled with a stochastic volatility described by the
Cox-Ingersoll-Ross equation. In the CIR process the volatility has
fluctuations of intensity $c$ and it reverts to a mean level $b$ at
rate $a$.

Our results show that as long as the mean level $a$ is different
from zero it is possible to observe a nonmonotonic behavior of MET
as a function of the two model parameters $b$ and $c$. The parameter
$a$ regulates the transition from a noise-only regime, where
reverting term is absent or negligible, to a revert-only regime,
where the noise term is absent or negligible. In the former case,
the enhancement of MET with a nonmonotonic behavior as a function of
the model parameters, that is the NES effect, is not observable. The
curves have a monotonic shape with a plateau. Moreover, if one of
the parameters is so big to push the system into a region where the
noise is greater than the barrier height of the effective potential,
the effect is no more observable at all. In the revert-only regime,
instead, the NES phenomenon is recovered. With its regulatory
effect, the reverting rate $a$ can be used to modulate the intensity
of the stabilizing effect of the noise observed by varying $b$ and
$c$. In this parameter region the probability density function of
the escape times of the returns fits very well that obtained from
the experimental data extracted by real market.

\section*{\label{sec:ack}Acknowledgements}
\vskip-0.48cm
 Authors wish to thank Dr. F.~Lillo for useful
discussions. This work was supported by MIUR, INFM-CNR and CNISM.

\end{document}